\documentclass[aps,twocolumn,prb,amsmath,amsfonts,showpacs]{revtex4}

\usepackage{graphicx}
\usepackage{bm}
\usepackage{color}
\begin{document}

\title{
Impact of off-diagonal exchange interactions on the Kitaev spin liquid state of $\alpha$-RuCl$_3$
}


\author{Daichi Takikawa}
\author{Satoshi Fujimoto}
\affiliation{Department of Materials Engineering Science, Osaka University, Toyonaka 560-8531, Japan }




\date{\today}

\begin{abstract}
Motivated by the recent experimental observation of the half-quantized thermal Hall conductivity for the candidate material of the Kitaev spin liquid, $\alpha$-RuCl$_3$[ Y. Kasahara {\it et al.}, Nature {\bf 559}, 227 (2018)], we investigate effects of non-Kitaev exchange interactions, which exist in the real material, on the gapped chiral spin liquid state realized in the case with an applied magnetic field. 
It is found that  off-diagonal exchange interactions 
enhance significantly the mass gap of Majorana fermions.
This result provides a possible explanation for robust quantization of the thermal Hall conductivity observed in the above-mentioned experiment. Furthermore, we demonstrate that if the Kitaev spin liquid state and the zigzag antiferromagnetic order coexist around the border of these two phases, off-diagonal exchange interactions can induce
Fermi surfaces of Majorana fermions, leading to a state similar to the U(1) spin liquid. 
\end{abstract}



\maketitle

\section{Introduction}
Quantum spin liquids (QSLs) are phases of magnetic materials with no conventional symmetry-breaking, but, instead, characterized by
a topologically nontrivial ground state with excitation gap, or gapless excitations emerging from
fractionalization of electron spins such as spinons.  
Exploration for spin liquid states in real magnetic materials is one of the most important long-standing issues of correlated electron systems.\cite{anderson}
The Kitaev honeycomb-lattice model is a two-dimensional (2D) exactly solvable model of a spin $s=1/2$ quantum magnet
which has a spin liquid ground state
with no symmetry-breaking.\cite{kitaev} 
A key feature of the Kitaev model is the existence of a strongly anisotropic exchange interaction 
(so-called the Kitaev interaction) which depends on the direction of bonds
on the honeycomb lattice, and suppress conventional magnetic long-range orders.
In the ground state of the Kitaev model, there is no magnetic correlation beyond the distance between neighboring sites,
and low-energy excitations are non-interacting gapless Majorana fermions.
When time-reversal symmetry is broken by an applied magnetic field $\bm{h}=(h_x,h_y,h_z)$, 
Majorana fermions acquire energy gaps $E_g\sim h_xh_yh_z/|K|^2$ with $K$ the Kitaev interaction, 
and the system turns into a chiral spin liquid state characterized by a nonzero Chern number, 
which possesses gapless chiral Majorana edge states on open boundaries. In this chiral spin liquid state, 
the thermal Hall conductivity is quantized as $\kappa=\frac{\pi k_{\rm B}^2T}{6\hbar}c$ with $c=\frac{1}{2}$ the central charge of
one-dimensional chiral Majorana fermions in edge states.
Further precise investigations have revealed 
transport, thermodynamics, and dynamical properties of the Kitaev model.\cite{PhysRevLett.112.207203,PhysRevB.92.115122,PhysRevLett.117.037209,PhysRevLett.119.127204,PhysRevLett.119.157203}

There are several candidate materials for the Kitaev honeycomb lattice model, in which the Kitaev interaction is realized
by strong spin-orbit couplings.\cite{Jackeli,PhysRevLett.105.027204,kitagawa,Katukuri_2014,Simon,1408.4811} 
Particularly, $\alpha$-RuCl$_3$ is promising for the realization of a QSL state.\cite{Banerjee,Yadav}  This system exhibits an antiferromagnetic (AFM) long range order with zigzag spin structures
at the temperature $T\sim 6$ K.\cite{PhysRevB.91.094422,PhysRevB.91.180401,PhysRevB.92.235119,Baek}   
The AFM order, however, is destroyed by applying a magnetic field parallel to honeycomb layers.
For the high field region, the realization of a spin liquid state has been suggested by extensive experimental studies such as neutron scattering measurements, NMR measurements, and specific heat measurement,\cite{Baek,Zheng,PhysRevB.91.180401,klanjsek} and also by subsequent theoretical studies.\cite{PhysRevB.97.241110}
Furthermore, the observation of the QSL phase under applied pressure was reported.\cite{Wang} 
Recently, quantization of the thermal Hall conductivity was observed in this putative QSL phase by Kyoto group,
which can be a strong evidence of the realization of the chiral spin liquid state.\cite{kasahara,kasahara2} 
According to this experimental result, the thermal Hall conductivity is half-quantized in agreement with the above-mentioned prediction
for temperatures below $\sim 5$ K, and for magnetic fields perpendicular to the honeycomb plane $h_{\perp} \sim  5 \sim 7$ T.
This remarkable observation has inspired several theoretical studies which particularly focus on the role of phonon excitations
for thermal transport.\cite{PhysRevLett.121.147201,PhysRevX.8.031032} 
An important but puzzling feature of this experimental result is that the temperatures for which the quantized thermal Hall conductivity is observed are almost the same order as
the energy scale of the Zeeman interaction due to the applied magnetic fields, which are believed to be an origin of
the energy gap $E_g$ of Majorana fermions in the Kitaev model.
Quantization of the thermal Hall conductivity is possible only for sufficiently low temperature regions $k_{\rm B}T \ll E_g$.
Thus,  the above experimental result suggests that the energy gap of Majorana fermions in the chiral spin liquid state of $\alpha$-RuCl$_3$ should be significantly larger than  $E_g\sim h_xh_yh_z/|K|^2$ which is estimated for
the ideal Kitaev model with the Zeeman term.

To understand the origin of the large energy gap of Majorana fermions, in this paper, we propose a scenario that the Majroana energy gap can be enhanced by non-Kitaev off-diagonal exchange interactions which exist in the real material $\alpha$-RuCl$_3$.
The off-diagonal exchange interactions arise from spin-orbit coupling, and according to {\it ab initio} studies, their magnitudes
are not negligible compared to the Kitaev interaction and the Zeeman energy due to 
applied magnetic fields.\cite{kee,PhysRevB.96.054434} 
There are two types of the off-diagonal exchange interactions, i.e. the $\Gamma$ term and the $\Gamma'$ term. (see eq.(\ref{eq:ham1})) 
Effects of the $\Gamma$ term, the magnitude of which is comparable to the Kitaev interaction,
have been extensively studied in previous studies.
In fact, the $\Gamma$ term is the origin of the zigzag AFM order of candidate materials for the Kitaev spin system.\cite{Rau,Katukuri_2014}
On the other hand, the $\Gamma'$ term, which is one order of magnitude smaller than the $\Gamma$ term,
has not been attracted much attention.
In this paper, we show that the $\Gamma'$ term combined with an external magnetic field can generate mass gap of Majorana fermions
which is substantially larger than that generated solely by the magnetic field.
Our findings explain qualitatively the origin of robust quantization of the thermal Hall effect against thermal excitations for
temperatures comparable to the energy scale of the Zeeman interaction, observed in the experiment.\cite{kasahara}

Our results also imply that the off-diagonal exchange interactions drastically affect the band structure of Majorana fermions
in the QSL state. Utilizing this property, we demonstrate that if the AFM zigzag order and the QSL state coexist around the border
of these two phases, Fermi surfaces of Majorana fermions with finite areas, which consist of
"electron"-like pockets and "hole"-like pockets can be realized. 
It is noteworthy that the Majorana Fermi surfaces have topological stability classified by the $K$ theory.\cite{horava,shiozaki,agterberg}
The realization of the Majorana Fermi surfaces
affects transport and thermodynamic properties of the Kitaev spin liquid in drastic ways.

The organization of this paper is as follows. In Sec. II, we discuss effects of off-diagonal exchange interactions on
the energy gap of Majorana fermions, and show that they enhance the Majorana gap significantly.
In Sec. III,  it is shown that Fermi surfaces of Majorana fermions are induced by a combined effect
of the off-diagonal exchange interactions and the coupling with magnetization due to the AFM order.
Summary is given in Sec. IV.

\section{Stabilization of the chiral spin liquid state due to off-diagonal exchange interactions}

\subsection{Enhanced energy gap of Majorana fermions due to off-diagonal exchange interactions}

\begin{figure}[h]
 \centering
   \includegraphics[width=6cm]{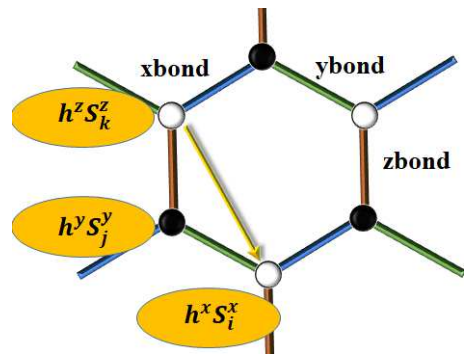}
 \caption{$x$,$y$,$z$-bonds on the honeycomb lattice. 
 The $A$ sub-lattice sites and the $B$ sub-lattice sites are, respectively, denoted as white and black circles.
 An example of the configuration of the sites $i$, $j$, and $k$ for the third-order perturbation term eq.(\ref{eq:third-h}) 
 }
 \label{fig:bond}
\end{figure}

We consider non-Kitaev interactions in addition to the Kitaev interaction which exist in the target material $\alpha$-RuCl$_3$.
According to an {\it ab initio} study,\cite{kee} the Hamiltonian for the 2D honeycomb-lattice plane in $\alpha$-RuCl$_3$ is given by,
\begin{eqnarray}
\mathcal{H}=\mathcal{H}_{J}+\mathcal{H}_{K}+\mathcal{H}_{\Gamma}+\mathcal{H}_{\Gamma'}, \label{eq:ham1}
\end{eqnarray}
\begin{eqnarray}
\mathcal{H}_{J}=J\sum_{ i,j  }\vec{S}_i\cdot\vec{S}_j, \label{eq:hamh}
\end{eqnarray}
\begin{eqnarray}
\mathcal{H}_{K}=-K\sum_{i,j \in \alpha{\rm -bonds}} S_i^{\alpha}S_j^{\alpha},  \label{eq:hamk}
\end{eqnarray} 
\begin{eqnarray}
\mathcal{H}_{\Gamma}=\Gamma\sum_{\substack{  i,j \in \alpha{\rm -bonds} \\ \beta,\gamma \neq \alpha}} [S_i^{\beta}S_j^{\gamma}+S_i^{\gamma}S_j^{\beta}], \label{eq:hamgam1}
\end{eqnarray}
\begin{eqnarray}
\mathcal{H}_{\Gamma'}=\Gamma'\sum_{\substack{  i,j \in \alpha{\rm -bonds} \\ \beta \neq \alpha}} [S_i^{\alpha}S_j^{\beta}+S_i^{\beta}S_j^{\alpha}],  \label{eq:hamgam2}
\end{eqnarray}
where $S^{\alpha}_i$ is an $\alpha=x,y,z$ component of an $s=1/2$ spin operator at a site $i$.
 $\mathcal{H}_{J}$ is the Heisenberg exchange interaction between the nearest neighbor sites, and
$\mathcal{H}_{K}$ is the Kitaev interaction between spins connected via $\alpha$-bonds with $\alpha=x,y,z$ (see FIG.\ref{fig:bond}).
$\mathcal{H}_{\Gamma}$ and $\mathcal{H}_{\Gamma'}$ are symmetric off-diagonal exchange interactions, the existence of which is generally allowed for edge-shared octahedra structures with strong spin-orbit couplings.
Similar {\it ab initio} Hamiltonian is also derived for another candidate material of the Kitaev spin system, 
Na$_2$IrO$_3$.\cite{Katukuri_2014,PhysRevB.96.054434}
$\alpha$-RuCl$_3$ exhibits the zigzag AFM order with the Neel temperature $T_N\sim 6$K.\cite{Banerjee,PhysRevB.91.094422,PhysRevB.91.180401,PhysRevB.92.235119,Baek} 
The magnetic long-range order is destroyed by an applied
magnetic field parallel to the honeycomb plane, which leads to the realization of a spin liquid state.\cite{Baek,Zheng}
Furthermore, as the applied magnetic field increases, quantization of the thermal Hall conductivity is observed, which implies the realization of the Kitaev spin liquid state with massive Majorana fermions.\cite{kasahara}
To focus on the Kitaev spin liquid state, we assume that the Kitaev interaction term $\mathcal{H}_{K}$ dominates the other terms, 
and deal with the other terms as perturbations.

The Kitaev Hamiltonian $\mathcal{H}_{K}$ can be diagonalized by using Majorana fermion representation of spin operators,
\begin{eqnarray}
S^{\alpha}_i=\frac{\rm i}{2}b^{\alpha}_ic_i,
\end{eqnarray}
where $b^{\alpha}_i$ and $c_i$ are Majorana fermion operators which satisfy the constraint condition,
$b_i^xb_i^yb_i^zc_i|\phi\rangle=|\phi\rangle$ with $|\phi\rangle$ the eigen state of the Kitaev spin liquid.
In terms of the Majorana fields, $\mathcal{H}_K$ is expressed as,
\begin{eqnarray}
\mathcal{H}_K=\frac{{\rm i}}{4}\sum_{i,j}\hat{A}_{ij}c_ic_j,   \label{eq:HK-ma}
\end{eqnarray}
where $\hat{A}_{ij}=\frac{1}{2}K\hat{u}_{ij}$ and $\hat{u}_{ij}={\rm i}b_i^{\alpha}b_j^{\alpha}$  with $i,j \in \alpha$-bond.
The $Z_2$ gauge fields $\hat{u}_{ij}$ are conserved, and can be replaced by the eigenvalues $\pm 1$. 
For $K>0$ ($K<0$), in the ground state, we can put $\hat{u}_{ij} \rightarrow 1$ ($-1$), and hence, $\hat{A}_{ij}\rightarrow \frac{1}{2}K$ ($-\frac{1}{2}K$).
Then, eq. (\ref{eq:HK-ma}) is reduced to the Hamiltonian of free massless Majorana fermions which can be diagonalized in the momentum representation.
When a magnetic field $\vec{h}=(h^x,h^y,h^z)$ satisfying $h^xh^yh^z\neq 0$
is applied to the system, the Zeeman interaction,
\begin{eqnarray}
\mathcal{H}_Z=-2\sum_{i,\alpha} S_i^{\alpha}h^{\alpha},
\end{eqnarray}
generates the mass-gap of Majorana fermions in the bulk, and a chiral spin liquid state with chiral gapless Majorana edge states
is realized. This chiral spin liquid state exhibits the quantum thermal Hall effect associated with a nonzero Chern number of the gapped Majorana fermion band.
The mass term is obtained by perturbative calculation up to the third order in $h^{\alpha}$, which leads to
three-spin interaction terms,
\begin{eqnarray}
\mathcal{H}^{(3)}\sim -\frac{8h^xh^yh^z}{K^2}\sum_{i,j,k}S^x_iS^y_jS^z_k, \label{eq:third-h}
\end{eqnarray}
where $i,j \in \alpha$-bond, and $j, k \in \beta$-bond with $\alpha\neq\beta$, as depicted in FIG.\ref{fig:bond}.
In  terms of Majorana fields, $\mathcal{H}^{(3)}$ is written into,
\begin{eqnarray}
\mathcal{H}^{(3)}\sim {\rm i}\frac{h^xh^yh^z}{K^2}\sum_{i,m}\sum_{p=1,2,3}(-1)^m c_{i+\bm{n}_p}c_i, \label{eq:mass1}
\end{eqnarray}
where $\bm{n}_1=(\frac{1}{2},\frac{\sqrt{3}}{2})$, $\bm{n}_2=(\frac{1}{2},-\frac{\sqrt{3}}{2})$, $\bm{n}_3=-\bm{n}_1-\bm{n}_2$,
and $m=0$ if the site $i$ is on the $A$ sub-lattice of the honeycomb lattice, and 
$m=1$ if $i$ is on the $B$ sub-lattice. (see FIG.\ref{fig:bond}).
This term yields
the Majorana mass gap $\Delta_0 \sim 4h^xh^yh^z/K^2$.  

\begin{figure}[h]
 \centering
   \includegraphics[width=6cm]{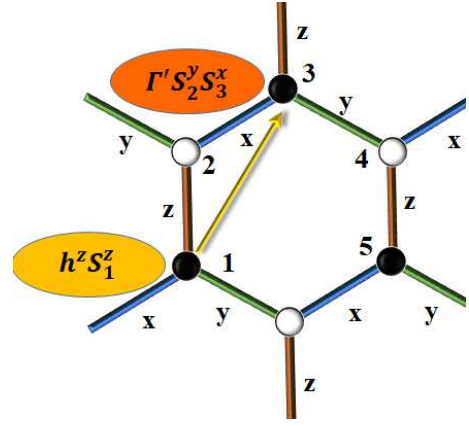}
 \caption{An example of $\Gamma'$ interactions which generate the mass term $\sim {\rm i}\frac{\Gamma'h^z}{K}c_3c_1$.}
 \label{fig:gammaprime}
\end{figure}

We, now, explore effects of the other terms in eq.(\ref{eq:ham1})  on the Kitaev spin liquid state.
It is noted that the Heisenberg interaction $\mathcal{H}_J$, and the off-diagonal exchange interactions 
$\mathcal{H}_{\Gamma}$, $\mathcal{H}_{\Gamma'}$ in eq.(\ref{eq:ham1}) do not
allow the representation in terms of Majorana fermions $c_i$ couplied with the $Z_2$ gauge fields $\hat{u}_{ij}$,
and also, these terms do not commute with  $\hat{u}_{ij}$; i.e. the $Z_2$ vortices (visons) are not conserved. 
Thus, we deal with these non-Kitaev interactions by using a perturbative calculation method.
Our perturbation calculation is based on the same idea as that of Kitaev’s paper for the derivation of the mass term due to magnetic fields,\cite{kitaev} and thus, it is valid provided that the energy scales of perturbations are smaller than the excitation gap of visons, which is of the order of the Kitaev interaction $\sim K$.
In the Kitaev spin liquid state which is the eigen state of visons, $\mathcal{H}_J$, $\mathcal{H}_{\Gamma}$, and $\mathcal{H}_{\Gamma'}$ are irrelevant perturbations,
provided that the magnitudes of the couplings $J$, $\Gamma$, and $\Gamma'$ are sufficiently small; i.e.
in the ground state of the Kitaev spin liquid with $\hat{u}_{ij}=1$ (or $-1$) for any $i$, $j$, the average of the interactions $\mathcal{H}_J$, $\mathcal{H}_{\Gamma}$, and 
$\mathcal{H}_{\Gamma'}$ are zero. 
The { \it ab initio} study for  $\alpha$-RuCl$_3$ implies that the magnitudes of $J$ and $\Gamma$ are comparable to or stronger than
the Kitaev interaction.\cite{kee}
For such large values of  $J$ and $\Gamma$,
magnetically ordered phases are more stabilized than the QSL state, as elucidated by mean field analysis and exact diagonalization studies.\cite{Rau}
However, the recent experimental studies revealed that the AFM order is suppressed 
by an applied in-plane magnetic field, leading to the transition to the QSL phase.\cite{Baek,Zheng,kasahara,kasahara2}
Our approach is based on the assumption that the Kitaev spin liquid state without $Z_2$ vortex is stabilized in this phase under the magnetic field. We investigate perturbative effects of the off-diagonal exchange interactions on the vortex-free spin liquid state.
The vortex-free Kitaev spin liquid state is stable as long as the energy scales of perturbations are sufficiently smaller than the excitation energy of the $Z_2$ vortex. We consider the parameter regime where this condition is satisfied, and it is legitimate to fix all the $Z_2$ gauge field $\hat{u}_{ij}=1$.
Even within this perturbation approach, it is found that the non-Kitaev interactions affect quantitative properties of the Kitaev spin liquid state in an important way.
In particular, 
$\mathcal{H}_{\Gamma'}$ combined with the Zeeman magnetic field can generate the mass gap up to the linear order in $h^{\alpha}$,
which can be much larger than the mass gap due to eq.(\ref{eq:mass1}) generated solely by magnetic fields.
Up to the second order in $\Gamma'$ and $h^{\alpha}$, we have obtained three-spin interaction terms  of the perturbed Hamiltonian which give the additional Majorana mass,
\begin{eqnarray}
\mathcal{H}^{(2)}_{\Gamma'}\sim \frac{2\Gamma'}{|K|}\sum_{\substack{  i,j \in \alpha{\rm -bonds} \\ j,k \in \gamma{\rm -bonds} \\ \beta \neq \alpha, \gamma}}
(h^{\alpha}+h^{\gamma})S^{\alpha}_iS^{\beta}_jS^{\gamma}_k \nonumber \\
\label{eq:pert2}
\end{eqnarray}
where $i$, $j$, and $k$ are arranged in the same way as eq. (\ref{eq:third-h}).
An example of the configuration of the $\Gamma'$ interaction and the Zeeman interaction is depicted in FIG.\ref{fig:gammaprime}.
In the ground state sector with no $Z_2$ vortex,
this perturbed Hamiltonian expressed in terms of Majorana fields is given by,
\begin{eqnarray}
\mathcal{H}^{(2)}_{\Gamma'}\sim -{\rm i}\frac{\Gamma'}{4|K|}\sum_{p,m}\sum_{\substack{  i,j \in \alpha{\rm -bonds} \\ j,k \in \beta{\rm -bonds} \\ i=k+\bm{n}_p}}(h^{\alpha}+h^{\beta})(-1)^m
c_ic_k, \label{eq:pert-gap}    
\end{eqnarray}
where $m=0$ ($m=1$) if the sites $i$, $k$ are on the $A$ ($B$) sub-lattice of the honeycomb lattice. 
Thus, this term linear in $h^{\alpha}$ gives the mass gap of Majorana fermions $\Delta_1\sim \Gamma'h^{\alpha}/(4|K|)$
in addition to $\Delta_0$ generated by eq.(\ref{eq:mass1}). 
We stress again that this result is valid provided that the energy scale of eq. (\ref{eq:pert-gap}), $\Gamma' h^{\alpha} / (4|K|)$,  
is sufficiently smaller than the vison gap $\sim K$.
The energy spectrum of Majorana fermions in this case is given by,
\begin{eqnarray}
E_{\pm}(\bm{k})=\pm \sqrt{|f(\bm{k})|^2+|\Delta(\bm{k})|^2},
\end{eqnarray}
with
\begin{eqnarray}
f(\bm{k})=\frac{K}{2}(e^{{\rm i}\bm{k}\cdot\bm{n}_1}+e^{-{\rm i}\bm{k}\cdot\bm{n}_2}+1),
\end{eqnarray}
\begin{eqnarray}
\Delta(\bm{k})&=&\Delta_0(\bm{k})+\Delta_1(\bm{k}),
\end{eqnarray}
\begin{eqnarray}
\Delta_0(\bm{k})&=&\frac{4h^xh^yh^z}{K^2}[\sin(\bm{k}\cdot\bm{n}_1) +\sin(\bm{k}\cdot\bm{n}_2)   \nonumber \\
&&+\sin(\bm{k}\cdot\bm{n}_3)],
\end{eqnarray}
\begin{eqnarray}
\Delta_1(\bm{k})&=&-\frac{\Gamma'}{|K|}[(h^x+h^z)\sin(\bm{k}\cdot\bm{n}_1) \nonumber \\
&&+(h^y+h^z)\sin(\bm{k}\cdot\bm{n}_2)   \nonumber \\
&&+(h^x+h^y)\sin(\bm{k}\cdot\bm{n}_3)].
\end{eqnarray}
The off-diagonal exchange interaction changes  the Majorana mass gap to $\Delta_0+\Delta_1$. 
In the case with $\Gamma'<0$, the magnitude of the energy gap of Majorana fermions is enhanced by the off-diagonal exchange interaction, which leads to robust stability of the chiral spin liquid state with the half-quantized thermal Hall conductivity.
In fact, this is indeed the case of $\alpha$-RuCl$_3$.\cite{kee}
Since the above analysis is based on perturbative expansion in $\Gamma'$ and $h^{\alpha}$, we can not
obtain quantitative estimates of the mass enhancement due to the $\Gamma'$ term. 
However, we would like to stress that this effect is substantial for material parameters of the candidate system  $\alpha$-RuCl$_3$, as will be discussed in the next subsection.
Furthermore, in contrast to eq.(\ref{eq:mass1}), which requires $h^xh^yh^z\neq 0$ for non-zero energy gap,
the mass-gap due to eq.(\ref{eq:pert-gap}) is finite unless $h^x+h^y+h^z = 0$ is satisfied.
 Thus, the gap is non-zero even when the magnetic field is parallel to one of the spin axes.
 
\begin{figure}[h]
 \centering
   \includegraphics[width=7cm]{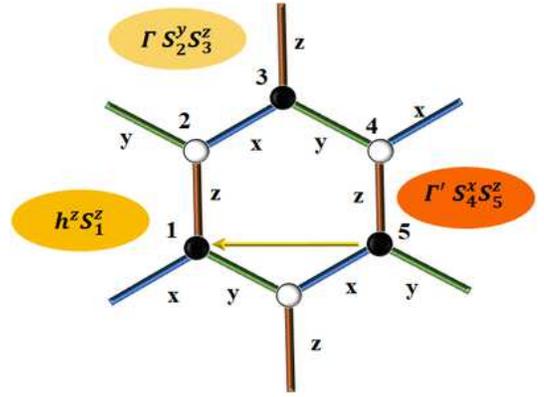}
 \caption{An example of $\Gamma$ and $\Gamma'$ interactions which generate the mass term eq.(\ref{eq:mass2}).}
 \label{fig:gamma}
\end{figure}

It is notable that the other terms in eq. (\ref{eq:ham1}) combined with a magnetic field can not generate the mass gap of Majorana fermions up to linear order in the magnetic field. Thus, although the magnitude of $\Gamma'$ is relatively smaller than
$K$ and $\Gamma$ for the case of $\alpha$-RuCl$_3$, its impact on the Majorana gap is crucially important.
Furthermore, the energy gap $\Delta_1(\bm{k})$  can be also enhanced by  $\mathcal{H}_{\Gamma}$.
The third order perturbative calculation in $h^{\alpha}$, $\Gamma'$, and $\Gamma$ results in the correction to the mass term eq.(\ref{eq:pert-gap})
given by,
\begin{eqnarray}
\mathcal{H}^{(3)}_{\Gamma'\Gamma} \sim {\rm i}  \frac{\Gamma'\Gamma}{K^2}\sum_{\gamma=x,y,z}\sum_{i,j} h^{\gamma}c_ic_k,
\label{eq:mass2}
\end{eqnarray}
where $i$ and $k$ are the next-nearest neighbor sites.
This term arises, for instance, from the combination of the interactions depicted in FIG.\ref{fig:gamma}

It is noted that  for the case of $\alpha$-RuCl$_3$, $\Gamma>0$,\cite{kee} and hence,
this correction term enhances the magnitude of the Majorana mass gap, leading to the robust stability of
the chiral spin liquid state characterized by the quantized thermal Hall effect.

The energy gap of itinerant Majorana fermions linear in a magnetic field was predicted before from a general symmetrical argument,\cite{PhysRevLett.117.037209} or by considering a model with the Dzyaloshinskii-Moriya interaction.\cite{LUNKIN2017} 
In contrast to these previous studies, however, our result is based on 
the microscopic analysis of the realistic model for candidate materials of Kitaev magnets.

It is noted that recent experimental studies for $\alpha$-RuCl$_3$ have verified the existence of the spin excitation gap linear in an applied magnetic,\cite{Baek,PhysRevLett.120.117204}  which is in agreement with our result obtained in this section.

\subsection{Implication for the experimental observation of the quantum thermal Hall effect in $\alpha$-RUCl$_3$}

According to the recent thermal transport measurement for $\alpha$-RuCl$_3$, the quantized thermal Hall conductivity,
$\kappa_{xy}=\frac{\pi k_B^2 T}{12\hbar}$, was observed for an applied magnetic field ranging from $4.5$ to $7$ Tesla (T) in the temperature region $4 \sim 5$ K; i.e. the Zeeman energy due to the applied magnetic field is almost the same order as $k_BT$.\cite{kasahara}
On the other hand, the realization of the quantization requires temperatures sufficiently lower than the energy gap of Majorana fermions. Thus, the experimental result implies that the energy gap of Majorana fermions should be an order of magnitude larger than that obtained from $\mathcal{H}^{(3)}$, i.e. $\Delta_0 \sim 4h_xh_yh_z/|K|^2$, the magnitude of which is much less than the Zeeman energy due to the magnetic field.
Our result of the Majorana energy gap due to the symmetric off-diagonal exchange interaction provides
a scenario that the enhanced Majorana gap $\Delta_0+\Delta_1$ stabilizes the quantization for the thermal Hall effect even for
temperatures comparable to the Zeeman energy due to applied fields.
In fact, according to the {\it ab initio} study,\cite{kee} $\Gamma'\sim -1.0\sim -0.5$ meV, and hence,
$\Delta_1$ contributes significantly the robust stability of the quantization of the thermal Hall conductivity.

\section{Field-induced Fermi surface of Majorana fermions in the coexisting phase of the spin liquid and the zig-zag antiferromagnetic order}

The result obtained in the previous section implies that internal magnetic fields caused by a magnetic order 
combined with the off-diagonal exchange interaction can drastically
change  energy band structures of Majorana fermions.
In this section, we demonstrate that this mechanism changes the Majorana Fermi point at zero energy into the Fermi surface of Majorana fermions with finite surface areas in the coexisting phase of the zigzag AFM order and the spin liquid state.
This scenario is of interest, because the density of states at the Fermi level is finite in this state, which leads to dramatic changes of thermodynamics and transport properties of the Kitaev spin liquid states. We discuss this possibility for the case of alpha-RuCl3, which exhibits the phase transition between the zigzag AFM phase to the QSL phase due to a magnetic field applied parallel to the honeycomb plane.\cite{Baek,Zheng,1807.06192} In fact, the character of the intermediate phase between the low-field zigzag AFM phase and the high-field fully spin-polarized phase has not yet been understood. Although the realization of the Kitaev spin liquid state is strongly suggested by experimental studies, the possibility of other quantum phases is not yet excluded. Thus, it is meaningful to examine various possible types of QSL phases.
If the phase transition from the zigzag AFM to the QSL phase is the first order, it is possible that honeycomb lattice layers with the zigzag AFM order and those with the spin liquid state may coexist.  Then, as shown in the next subsection, a gapless spin liquid phase with the Majorana Fermi surface can be realized in the border between the AFM phase and the chiral spin liquid phase. 

There are some preceding studies on the realization of Majorana Fermi surfaces in the Kitaev model; e.g.
cases of a decorated square lattice and a three-dimensional hyperoctagon lattice.\cite{0908.1614,PhysRevB.89.235102}
To this date, however, there are no real candidate materials
with these lattice structures in which the Kitaev interaction is sufficiently strong. 
Thus, the result presented in this section provides a more realistic scenario for the realization of Majorana Fermi surfaces.

Recently, the realization of a U(1) spin liquid state with the spinon Fermi surface in the intermediate phase between the zigzag AFM ordered phase for low magnetic fields and a spin-polarized phase for high magnetic fields are proposed 
by several groups.\cite{Liujun,Niravkumar,Yi-Fan,Hickey}
The origin of the Fermi surface of Majorana fermions found here is quite different from these previous proposals.

\subsection{Field-induced Fermi surface of Majorana fermions}

We, first, clarify necessary conditions for the realization of the Majorana Fermi surface.
Since the Hamiltonian of the Kitaev model with the mass term, $\mathcal{H}_K+\mathcal{H}^{(2)}$, 
expressed in terms of Majorana fields possesses particle-hole symmetry, the emergence of the finite Fermi surface areas of the non-degenerate Majorana bands requires broken inversion symmetry. 
A simple way of breaking inversion symmetry in the Kitaev spin liquid state is to introduce an internal magnetic field
due to a conventional collinear AFM order on the honeycomb lattice; i.e. spins on two sub-lattice of the honeycomb lattice are
aligned in the opposite directions. 
We can easily verify that if the layer with the AFM order and the layer in the QSL state coexist, and are coupled via the exchange interaction,
the internal field combined with the $\Gamma'$ term gives rise to the energy shift of the Majorana cone band, 
leading to finite areas of Majorana Fermi surfaces.
However, in the case of $\alpha$-RuCl$_3$, the AFM order has the zigzag structure, and this simple scenario is not applicable.
In the case that the zigzag AFM ordered layer is coupled with the QSL layer, 
there are four inequivalent sites in a unit cell of the honeycomb lattice in the QSL layer. 
We examined the band structure of Majorana fermions in this coexistence phase numerically, and found that Majorana fermi surfaces do not appear.
Thus, we consider a bit more complicated situation with multi-layers of honeycomb-lattice phases in which the zigzag AFM orders
have different zigzag directions, which strengthens breaking of inversion symmetry.

\begin{figure}[h]
 \centering
   \includegraphics[width=9.cm]{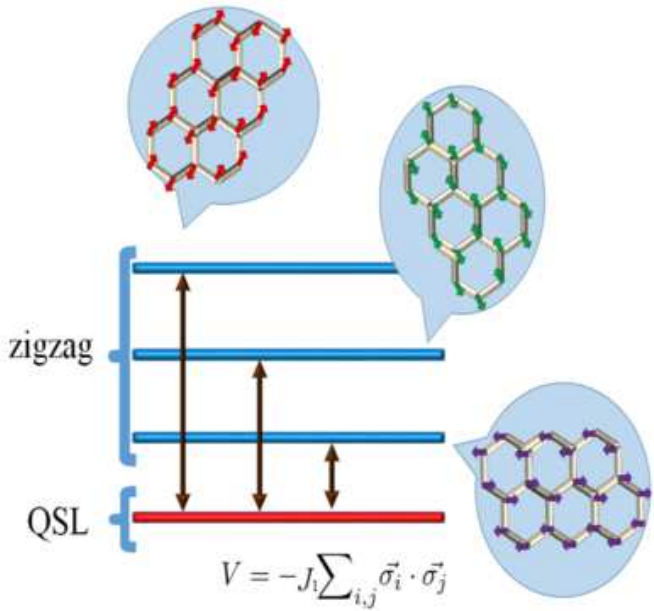}
 \caption{stacked layer structure composed of the QSL phase and the zigzag AFM order phases with three different zigzag directions
  $(1,0)$, $(-\frac{1}{2},\frac{\sqrt{3}}{2})$, $(-\frac{1}{2},-\frac{\sqrt{3}}{2})$ on the honeycomb lattice plane.}
 \label{fig:kyoukaisou}
\end{figure}

\begin{figure}[h]
 \centering
   \includegraphics[width=7cm]{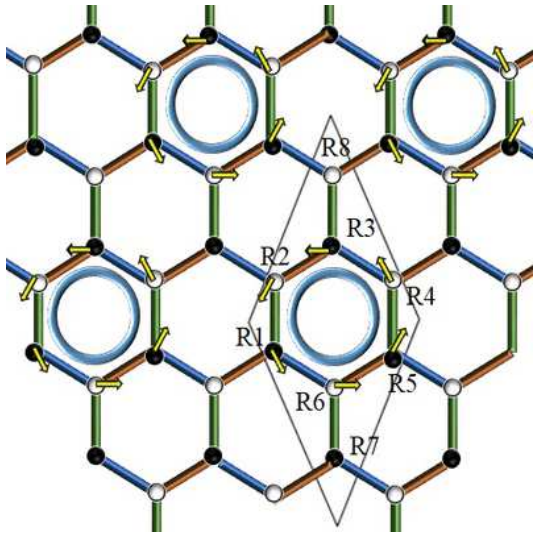}
 \caption{Internal magnetic field due to the exchange interaction between the three zigzag AFM layers and the QSL layer.}
 \label{fig:jikisou3}
\end{figure}

\begin{figure}[h]
 \centering
   \includegraphics[width=8.5cm]{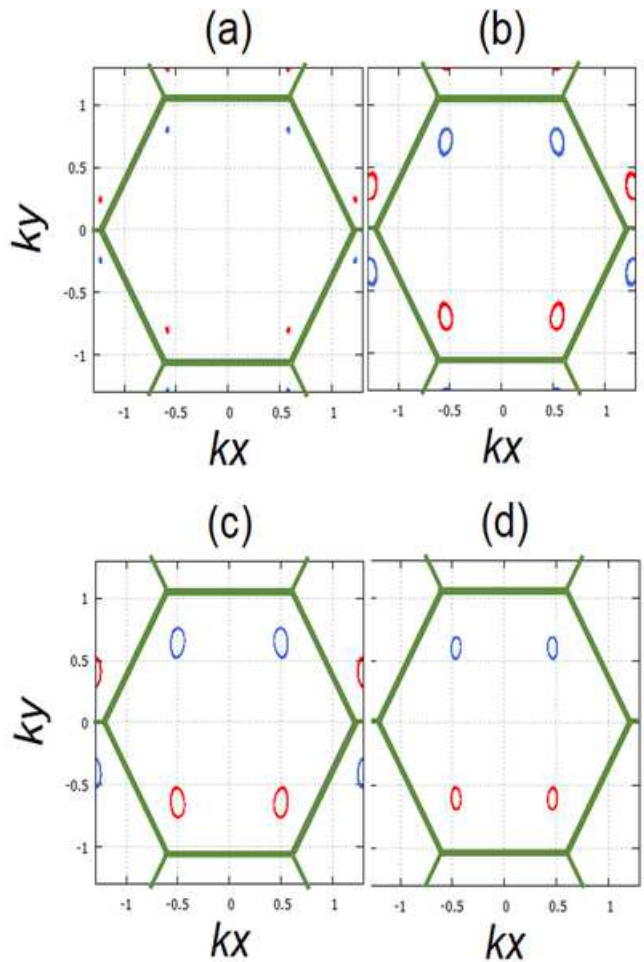}
 \caption{Majorana Fermi surfaces for a magnetic field $\bm{h}=(h,0)$ with (a) $h=2|K|$, (b) $h=3|K|$,
 (c) $h=4|K|$,  (d) $h=5|K|$.  $\Gamma'=0.148K$.
 "Electron"-like ("hole"-like) Fermi surfaces are depicted in blue (red) color.
 }
 \label{fig:FS}
\end{figure}

\begin{figure}[h]
 \centering
   \includegraphics[width=8cm]{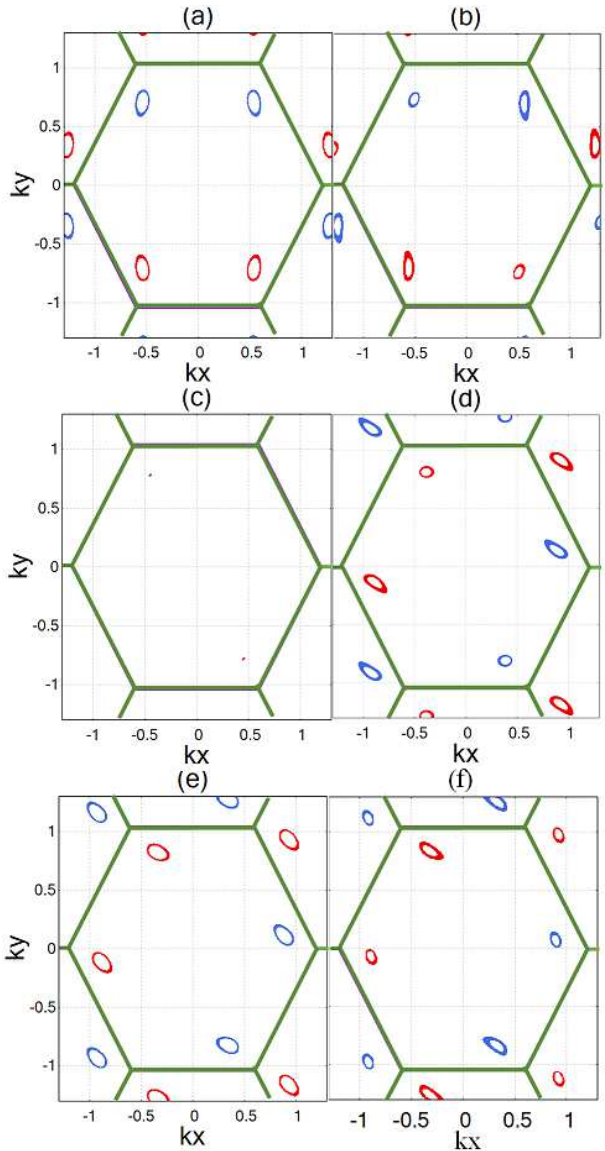}
 \caption{Majorana Fermi surfaces for a magnetic field $\bm{h}=h(\cos\theta,\sin\theta)$ with $h=3|K|$, $\Gamma'=0.148K$,
and (a) $\theta=0$, (b) $\theta=\pi/12$,
 (c) $\theta=\pi/6$,  (d) $\theta=\pi/4$, (e)  $\theta=\pi/3$, (f) $\theta=5\pi/12$. "Electron"-like ("hole"-like) Fermi surfaces are depicted in blue (red) color.
 }
 \label{fig:MFS}
\end{figure}

To be concrete, we consider a multi-layer structure composed of a QSL layer and three zigzag AFM layers with different zigzag directions along $(1,0)$, $(-\frac{1}{2},\frac{\sqrt{3}}{2})$, $(-\frac{1}{2},-\frac{\sqrt{3}}{2})$ on the honeycomb lattice planes, 
as depicted in FIG. \ref{fig:kyoukaisou}.\cite{com1}
To simplify the analysis, we assume that the exchange coupling between these ordered layers and the QSL layer are
 the same in the magnitude. This simplification is not essential for the realization of the Fermi surfaces, because, as will be elucidated below, the Majorana Fermi surfaces in the Kitaev spin liquid state has topological stability. 
 The effective exchange fields acting on spins in the QSL layer have  the spatial structure as shown in FIG.\ref{fig:jikisou3}.
We also add a uniform magnetic field $\bm{h}=h(\cos\theta,\sin\theta)$ applied parallel to the honeycomb lattice plane, where
$\theta$ parametrizes the direction of the magnetic field on the honeycomb lattice plane.\cite{com1}
It is found that Majorana Fermi surfaces appear in this case for sufficiently large couplings with the exchange fields.
In FIG.\ref{fig:FS}, we show the calculated results of Fermi surfaces of Majorana fermions in the case with the exchange field $h_{\rm ex}=10|K|$ and the external magnetic field $h=2|K|, 3|K|, 4|K|$ and $5|K|$ applied parallel to $(1,0)$ direction.
We used rather large values of $h_{ex}$ and $h$ to make the size of the Fermi surfaces large enough to be visible. 
The origin of the Fermi surfaces is the second order perturbation term, eq. (\ref{eq:pert-gap}).
We stress that the energy scale of this term, $\Gamma' h/ (4|K|)$ or $\Gamma' h_{ex} / (4|K|)$, is still smaller than the vison gap $\sim  K$ even for such large values of $h$ and $h_{ex}$, since the magnitude of $\Gamma'$ is set to a small value, 
$\Gamma'= 0.148 |K|$. Thus, our perturbation approach is valid for parameters used in these calculations.
The Fermi surfaces consist of "electron"-like pockets and "hole"-like pockets.
The shape of the Fermi surface notably depends on the direction of the applied magnetic field.
In FIG.\ref{fig:MFS}, we show the dependence of the Fermi surface on the magnetic field angle.
Since the spatial structure of the exchange field breaks inversion symmetry as well as time-reversal symmetry,
the asymmetric Fermi surfaces appear.
It is noted that the Hamiltonian with multi-species of Majorana fermions for the multi-layer system is 
diagonalized by a unitary transformation, and hence, the Fermi surfaces are formed by complex fermions arising 
from the unitary transformation of the Majorana fermions.
However, there is an important difference between our system and conventional complex fermion systems.
Since our Hamiltonian possesses particle-hole symmetry, a complex fermion with a momentum $\bm{k}$ and an energy
$E_{\bm{k}}$ and that with a momentum $-\bm{k}$ and an energy $-E_{\bm{k}}$ are not independent. The anti-commutation relation between these two fermions is nonzero.
Nevertheless, as long as low-energy properties are concerned, and any interactions between these two fermions are neglected,
the state with the Fermi surfaces can be regarded as a complex fermion system with U(1) symmetry.
More precisely, there are eight inequivalent sites in the unit cell on the spin-liquid layer as depicted in FIG. \ref{fig:jikisou3}. As a result, there are eight species of Majorana fermions, which lead to four bands in the Brillouin zone as shown in FIGs. \ref{fig:FS} and \ref{fig:MFS}. These Fermi surfaces are formed by the linear combination of the eight Majorana fermions with complex-number coefficients. Thus, they have the U(1)  degrees of freedom.
This state is similar to the U(1) spinon spin liquid discussed recently as a promising candidate of
the intermediate phase between the zigzag AFM phase and a fully-spin-polarized phase.\cite{Liujun,Niravkumar,Yi-Fan}

We would like to stress that our scenario for the realization of the Majorana Fermi surfaces is not restricted to the specific setup shown in FIG. \ref{fig:kyoukaisou}, but it is potentially 
applicable to any candidate materials for Kitaev magnets provided that  honeycomb layers with an AFM order which breaks inversion symmetry coexist with layers in the Kitaev spin liquid state.

\subsection{Topological stability of Majorana Fermi surface}

It is known that there are universality classes of stable Fermi surfaces which are classified by $K$ theory, and protected against local perturbations by topological invariants.\cite{horava,shiozaki,agterberg}
Fermi surfaces in two dimensions can be classified by the homotopy group $\pi_0(\mathcal{H})$ with $\mathcal{H}$ the Hilbert space
of the system; i.e. Fermi surfaces divide the momentum space into disconnected regions where a topological invariant takes different values, ensuring topological stability.
In this argument, the stability of gapless states at each $\bm{k}$-point on the Fermi surface is considered, and thus, symmetry operations which do not  change the position of momentum $\bm{k}$ are relevant.
For the case of the Majorana Fermi surface found in the previous section, the  system has neither time-reversal symmetry nor inversion symmetry, and is classified as class A with no symmetry. Note that the particle-hole symmetry of the Kitaev Hamiltonian does not play any roles because of the above-mentioned reason. 
For class A, $\pi_0(\mathcal{H})=\bm{Z}$. Thus, the Majorana Fermi surface is protected by the $\bm{Z}$ invariant which is nothing but the total number of Fermi surfaces.

\section{Summary}

In this paper, effects of off-diagonal exchange interactions on the Kitaev spin liquid state is investigated.
It is found that the off-diagonal exchange interactions can enhance significantly the magnitude of the mass gap of Majorana fermions
generated by applied magnetic fields. This effect provides possible explanation of the robust quantization
of the thermal Hall conductivity experimentally observed by Kyoto group.\cite{kasahara}

We have also revealed that in the vicinity of the phase boundary between the zigzag AFM phase and the QSL phase
under applied magnetic fields, 
the Fermi surface of Majorana fermions can be realized. This phase is similar to
the U(1) spinon spin liquid, since the Fermi surface is formed by complex fermions 
obeying usual anti-commutation relations, 
as long as low-energy states in the vicinity of the Fermi surface
is concerned.

It is noted that the diagonal and off-diagonal exchange interactions also generate mutual interactions between the complex fermions,
which may induce the Fermi surface instability.\cite{PhysRevLett.115.177205}
It is an interesting future issue to investigate possible phase transitions such as spin-Peierls instability
of the Fermi surfaces.

\begin{acknowledgments}
The authors are grateful to Y. Kasahara, M. Klanjsek, Y. Matsuda, T. Mizushima, K. Shiozaki and A. Tsuruta for valuable discussions. 
 This work was supported by the Grant-in-Aids for Scientific
Research from MEXT of Japan [Grants No. 17K05517, and KAKENHI on Innovative Areas ``Topological Materials Science'' [No.~JP15H05852]  and "J-Physics" [No.~JP18H04318]. 
\end{acknowledgments}

\bibliography{kitaev-ref}
\bibliographystyle{apsrev}


\end{document}